\documentclass[preprint,flushrt,eqsecnum]{aastex}

\slugcomment{To be submitted to ApJ}

\shortauthors{Wren et al.}
\shorttitle{Observations of the Optical Counterpart to XTE J1118+480}

\begin{document}

\title{Observations of the Optical Counterpart to XTE J1118+480 During
Outburst by the ROTSE-I Telescope}

\author{J. Wren\altaffilmark{1},C. Akerlof\altaffilmark{2},
R. Balsano\altaffilmark{1}, J. Bloch\altaffilmark{1},
K. Borozdin\altaffilmark{1},
D. Casperson\altaffilmark{1}, G. Gisler\altaffilmark{1},\\
R. Kehoe\altaffilmark{2,4}, B. C. Lee\altaffilmark{2,4},
S. Marshall\altaffilmark{3}, T. McKay\altaffilmark{2}, 
W. Priedhorsky\altaffilmark{1}, 
E. Rykoff\altaffilmark{2}, D. Smith\altaffilmark{2},\\
S. Trudolyubov\altaffilmark{1}, W. T. Vestrand\altaffilmark{1}}
\email{e-mail:jwren@lanl.gov}

\altaffiltext{1}{Los Alamos National Laboratory, Los Alamos, New Mexico 87545}
\altaffiltext{2}{University of Michigan, Ann Arbor, Michigan 48109}
\altaffiltext{3}{Lawrence Livermore National Laboratory, Livermore, California 94550}
\altaffiltext{4}{Fermi National Accelerator Laboratory, Batavia, Illinois 60510}

\begin{abstract}

The X-ray nova XTE~J1118+480 exhibited two outbursts in the
early part of 2000.  As detected by the Rossi X-ray Timing Explorer (RXTE),
the first outburst began in early January and the second began in
early March.  Routine imaging of the northern sky by the Robotic
Optical Transient Search Experiment (ROTSE) shows the optical
counterpart to XTE~J1118+480 during both outbursts.  These data
include over 60 epochs from January to June 2000.  A search of
the ROTSE data archives reveal no previous optical outbursts of
this source in selected data between April 1998 and January 2000.  
While the X-ray to optical flux ratio of XTE~J1118+480 was low
during both outbursts, we suggest that they were full X-ray nov{\ae}
and not mini-outbursts based on comparison with similar sources.
The ROTSE measurements taken during the March 2000 outburst also
indicate a rapid rise in the optical flux that preceded the X-ray
emission measured by the RXTE by approximately 10 days.  Using these
results, we estimate a pre-outburst accretion disk inner truncation
radius of $\sim 1.2 \times 10^{4}$ Schwarzschild radii. 
~\\

\end{abstract}

\section{Introduction}
\renewcommand{\thefootnote}{\fnsymbol{footnote}}
\setcounter{footnote}{0}

On March 29, 2000, the detection of a new X-ray nova
with the RXTE All Sky Monitor (ASM) was announced \citep{Remillard}.
Subsequent analysis of archival ASM data revealed that the
source had been rising since March 5 and had also undergone an earlier
outburst during January 2-29, 2000.  An optical counterpart was
quickly identified at 12.9 magnitude by \citet{Uemura}, who also
identified an 18.8 magnitude object from the USNO catalog as the
probable quiescent optical counterpart.
Observations with the Burst and Transient Source Experiment
(BATSE) indicate that the energy spectra for both outbursts was
characterized by a power law with a photon index of 2.1, and the
source was visible up to 120 keV \citep{Wilson}.
The optical spectrum was typical of X-ray nov{\ae} in outburst
\citep{Garcia}.  Optical data also revealed a sinusoidal variation
suggesting a binary system with an orbital period of 4.1 hours
\citep{Cook}.  Quasi-periodic oscillations (QPOs) of 0.08 Hz were found
using the RXTE Proportional Counter Array (PCA) \citep{RSB00}.
No periodic signal was observed in the power spectrum at
frequencies greater than 100 Hz.  The above measurements all imply
that XTE J1118+480 is a binary accretion system composed of a black
hole and an evolved low-mass companion.  Recently, observations of
the system in quiescence by \citet{McClintock01} have established that
the primary of the system has a mass greater than $6.00 \pm 0.36~M_\odot$,
thus indicating that it is a very strong black hole candidate.

XTE~J1118+480 exhibits some features that are atypical of X-ray
nov{\ae}.  The X-ray to optical flux ratio is very low
which would generally indicate a high inclination system, however
no eclipses have been observed \citep{Uemura}.
The source is near the Lockman hole
\citep{Lockman} and does not suffer from high interstellar
absorption, thereby allowing observations with the Extreme Ultraviolet
Explorer (EUVE).  The EUVE observations show no periodic
modulation suggesting the inclination is low enough that no
obscuration by the disk rim occurs \citep{Hynes00}.  This implies that
the source is intrinsically faint in X-rays.

Due to the strange behavior of the source, optical data during
the initial stages of these outbursts provide useful constraints on the
possible structure of the accretion disk and the mechanisms involved.
The Robotic Optical Transient Search Experiment I (ROTSE-I) imaged the
source throughout both the January and March outbursts.  These data
are presented below.

\section{Instrument and Observations}

The ROTSE program consists of several robotic telescopes designed to
search for optical transients, particularly those associated with
Gamma-Ray Bursts (GRBs) (Kehoe et al. 2000).  These robotic
telescopes rapidly respond to satellite derived triggers of
astrophysical transients in various wavebands.  The first generation
of the ROTSE telescopes is ROTSE-I, which consists of an array of four
Canon 200mm f1.8 lenses, each equipped with a thermo-electrically
cooled 2048x2080 pixel CCD camera.  Each lens/camera pair has an
8.1x8.1 degree field.  Each image pixel subtends 14.4 arcseconds.
To maximize sensitivity, the system is currently operated without
filters.  The system allows each camera to expose simultaneously so
that the array acts as a single telescope with a 16x16 degree field.

When not responding to triggered events, the ROTSE telescopes perform
regular patrol observations.  A typical ROTSE-I sky-patrol consists of
a series of 80 second exposures covering the area of sky visible above
20 degrees elevation. Each field is imaged twice consecutively to
allow for the elimination of false detections caused by cosmic rays,
satellite trails and glints, hot pixels, etc.  The mount is moved
slightly between exposure pairs to better facilitate the background rejection
by shifting the celestial coordinates of hot pixels and other camera
defects.  Currently we perform two sky-patrols every night, generating
four images (two pairs) of most sky locations.   During periods near
full moon, the exposure time is reduced to 20 seconds to eliminate the
possibility of saturation due to the increased sky brightness. 

The next morning, the images ($\sim$1000 total) are corrected and
calibrated to obtain accurate positions and magnitudes for each object
by comparison to the Hipparchos catalog \citep{Hog}.  This provides
accurate positions and an effective V-band magnitude for each object.
Normal sky-patrol images are generally sensitive to 15th magnitude
(14th during full moon).  A typical night's observing covers about
18,000 square degrees and records the brightness of $\sim 10^7$ stars.
For objects within 2 degrees of XTE~J1118+480, over 22 observations
during the month of March 2000, photometric accuracy was $\pm0.02$ at
magnitude 13 and $\pm0.14$ at magnitude 15.
Astrometric residuals are generally about 0.1 pixel (1.4 arcseconds).
ROTSE-I has been operating in this manner since March of 1998.  The
archive has now grown to over 4 Terabytes ($\sim$660,000 images)
covering the entire sky north of -40 degrees declination.

\section{Results}

\begin{figure}[tb]
   \epsscale{1.0}
   \plotone{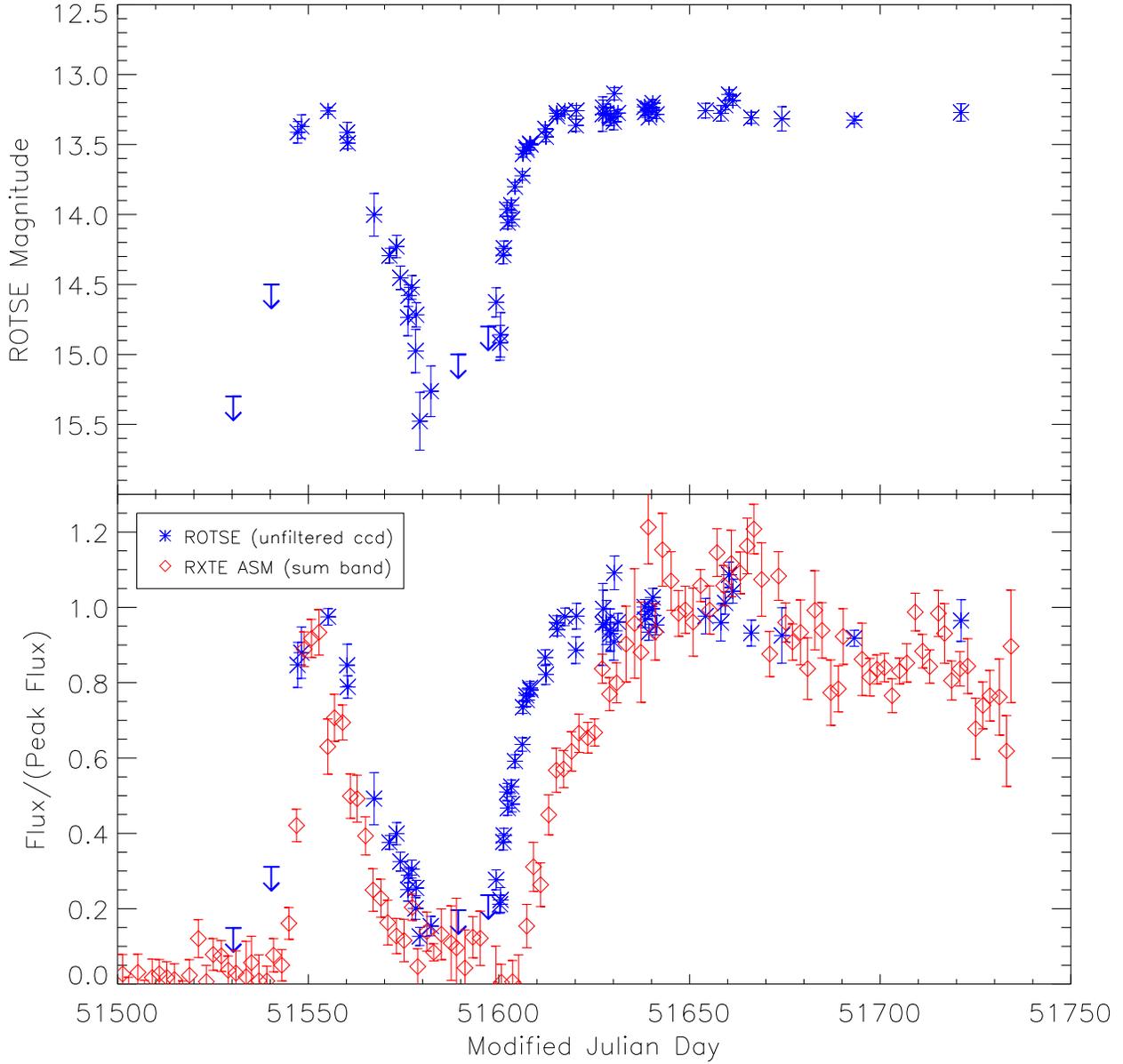}
   \caption{The top plot shows the ROTSE-I light-curve for XTE
            J1118+480.  The lower plot is a comparison
            of the ROTSE-I and RXTE/ASM fluxes over the same time
            period.  The ASM fluxes are 2 day averages.
            The peak fluxes used for the scaling were
	    $1.85\times10^{-14} erg/s/cm^{2}/\textrm{\AA}$ for ROTSE and 
            $2.8 c/s$ for the ASM.}
   \label{fig:flux}
\end{figure}

The ROTSE-I archive provides data on the optical counterpart
to XTE~J1118+480 throughout the January outburst and most of the 
March 2000 outburst (Figure \ref{fig:flux}).  The ROTSE-I system was
inactive from December 30, 1999, to January 4, 2000, in preparation
for the year 2000 roll-over.  As a result, we do not have data on the
initial stages of the January outburst.  The object is especially well
covered from mid-February through March.  After mid-April, our
coverage dropped significantly and we did not observe the final stages
of the second outburst.  During the duration of our
observations of this source, we detected no short optical flares as
mentioned by \citet{Uemura}.  For the analysis presented here, the
data was normalized to the observation from March 1 to ensure proper
relative photometry between images.  Also, since XTE~J1118+480 was
relatively close to our detection threshold in some of the shorter
exposures, we took the weighted average of each pair of sky patrol
observations.  A search of the ROTSE-I archives during periods of new
moon going back to April 1998 reveal no indication of this source.
The location of XTE~J1118+480 is typically not observed by ROTSE-I
during the fall months of the year ($\sim$ August-October).  

\begin{figure}[tb]
   \plotone{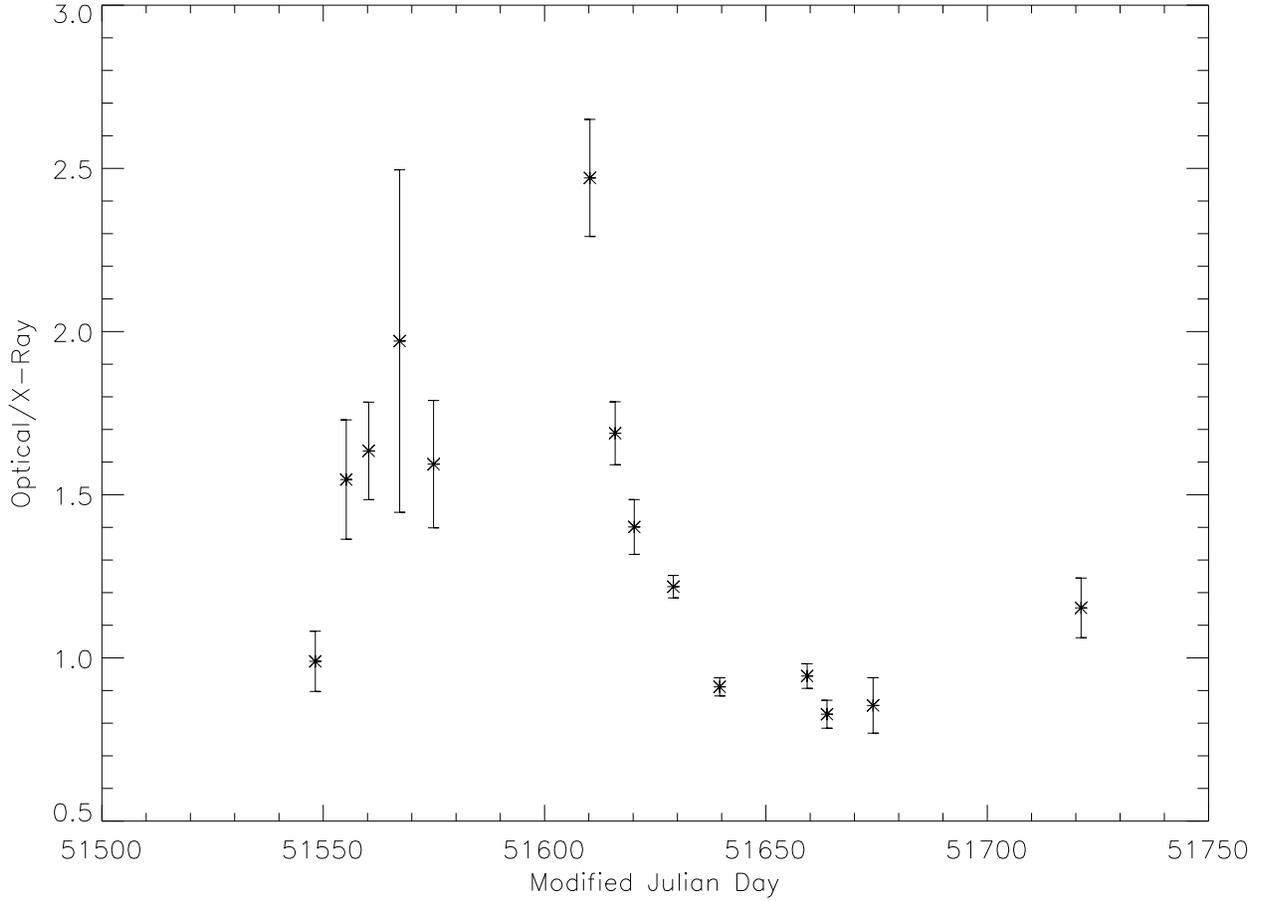}
   \caption{The ratio of the flux observed by ROTSE-I and the
   RXTE/ASM.  Values greater than 1 during the declining phase
   indicate the optical emission is lagging the X-ray emission.
   Values greater than 1 during the rising phase indicate the
   optical emission is leading the X-ray emission.  Fluxes were scaled
   as in figure \ref{fig:flux}.  The ASM values used were a weighted
   average of the values within 2 days of each ROTSE-I measurement.
   Resulting ratios were then binned in 6 day increments.  The ratios
   which correspond to ASM values that were consistent with zero are
   not shown.}
   \label{fig:hardratio}
\end{figure}

The ROTSE-I light curve from the January burst has a similar
morphology to the RXTE/ASM light curve.  Both display a rapid rise which
peaked around January 6.  The X-ray data indicate that the object reached
peak flux in less than 10 days.  Although the initial
stages of the January outburst were poorly sampled by ROTSE-I, the
early limits indicate that the optical rise must have been quite rapid
as well.  The object then proceeded to dim at a relatively linear rate
in flux over the next month in both the optical and X-ray.  The
later stages of the decay were well monitored by ROTSE-I.  Figures
\ref{fig:flux} and \ref{fig:hardratio} suggest that the optical
emission lagged the X-ray emission during the declining phase of this
burst by $\sim$ 5 days.  

ROTSE-I detected the second outburst of XTE~J1118+480 on February
26, 2000, 7 to 8 days before the first significant detections by the
ASM.  The optical intensity rapidly rose to half its peak intensity
by March 1 and achieved a peak intensity of magnitude 13.3, five
days later.  In contrast, the X-ray emission reached half its peak
intensity by March 11, approximately 10 days after the optical
reached half its peak intensity level.
XTE~J1118+480 remained near magnitude 13.3 through June 2000.  The
January outburst also peaked near magnitude 13.3, possibly
indicating that this is a natural limit based on the structure of the
accretion disk.  Figure \ref{fig:hardratio} shows the optical to X-ray
flux ratios seen by ROTSE-I and the ASM respectively.  During the
early stages of March burst, the flux ratio initially was high
and then decreased steadily to unity over the next 20 days or so.

\section{Discussion}

Optical precursors have been observed previously in black
hole X-ray nov{\ae}.  Just prior to an X-ray outburst of GRO J1655--40
in April 1996, \citet{Orosz} obtained filtered optical data
on the source showing an optical rise preceding the X-ray outburst
by about 6 days.  Optical observations of the August 1993
mini-outburst of GRO J0422+32 also indicated optical emission
preceding the detection of X-rays by 16 days \citep{Alberto}.
\citet{Hynes00} have suggested the outbursts of XTE~J1118+480 are more
like mini-outbursts seen in GRO J0422+32 than a full X-ray nova.
The X-ray delay observed in the March outburst of XTE~J1118+480 is
also analogous to the well known UV delay observed for dwarf nov{\ae}
(\citet{Warner} and references therein).

While the peak X-ray luminosities for both outbursts of XTE~J1118+480
($\sim 10^{36}$ erg/s in the $3 - 100$ keV energy 
range, assuming the source distance of 1.1 kpc) were rather low, the
general properties of its long-term variability are similar
to that of other bright X-ray transients.  For example, XTE~J1550--564
also underwent two consecutive outbursts of comparable duration to
XTE~J1118+480 beginning in late 1998 \citep{Jain01}.  Moreover, the
normalized profiles of the second outbursts in both systems are very
similar. This fact suggests that this outburst of XTE~J1118+480 is a
full X-ray nova rather than a mini-outburst. The low
peak luminosity of the source can probably be attributed to the
properties of the binary system. The binary period of XTE~J1118+480 is
known to be one of the shortest among the low mass X-ray binary
transient systems, implying a smaller size of the 
primary Roche lobe and different parameters of the accretion disk (like mass 
and accretion rate). The X-ray and optical properties of XTE~J1118+480 
\citep{RSB00,Hynes00} resemble the properties of another well-known 
Galactic black hole candidate GX~339--4 during its hard/low state outbursts 
(at comparable luminosity level of $10^{35} - 10^{37}$ ergs/s) 
\citep{Motch83}. Both sources demonstrate a high optical to X-ray flux ratio 
and simultaneous low frequency QPOs in X-ray and optical bands
\citep{Motch83,RSB00,Hynes00}. 

To explain the origin of dwarf nov{\ae} and X-ray transients, the
Disk Instability Model (DIM) was proposed \citep{Smak81,MW89}. In the
framework of this model, the UV/X-ray delay is a result of propagation
of the heating front through the accretion disk \citep{Meyer84}. This
front transforms the disk from the cold (quiescent) state to a hot state
raising the optical and UV/X-ray flux from the disk. However, the
standard DIM fails to explain the delay quantitatively: the calculated
travel time of the heating front, which should be near the sound speed
of the medium, is less than 1 day for a typical system -- a value
which is much shorter than commonly observed \citep{PVW86,Hameury97}.
 
A number of accretion flow models have been proposed to explain
the hard X-ray emission of X-ray binaries and cataclysmic variables 
in the quiescent and hard/low spectral states
\citep{CT95,MMH94,NMY96}.  These models involve a hot optically-thin
inner region that is surrounded by an optically-thick standard
accretion disk. This two-component geometry of the accretion flow
seems to resolve the problem of the standard DIM. The inward moving
heating front should stop at the inner edge of the optically-thick
disk.  Then the inner edge of the transformed disk moves toward the
compact object on the viscous time scale, which is much longer than
the heating front propagation time.  As the transition radius moves
inward, a growing fraction of the emitted photons are intercepted
and Comptonized by the hot corona, giving rise to the hard X-ray
emission.  The X-ray delay times predicted by this model are in much
better agreement with observations.

One proposed model involving a hot optically-thin inner region of the
accretion disk is the Advection Dominated Accretion Flow or ADAF
\citep{NMY96}.  In this model the accretion disk evaporates into an
optically-thin, quasi-spherical corona.  The corona, or ADAF region,
consists of a two-temperature plasma in which the ions and electrons
interact weakly which results in much of viscous heat being
advected into the black hole instead of being radiated away through
Comptonization.  \citet{Esin01} have applied the ADAF model to
multiwavelength observations of XTE~J1118+480 during outburst in
optical, EUV, and X-ray.  These multiwavelength measurements are well
fit by a $\sim 9 M_\odot$ black hole at a distance of $\sim 1.1$ kpc
in which the ADAF transition radius, $r^{tr}$, is at $55 R_{s}$ where
$R_{s}$ is a Schwarzschild radius.

\citet{Hameury97} use the ADAF model and the results of \citet{Orosz}
to estimate the transition radius of the quiescent
optically thick accretion disk in GRO~J1655--40 to be on the order
of $\sim 10^{4} R_{s}$.  Assuming the same mechanism for the outburst
of XTE~J1118+480 and given the X-ray delay time observed in the March
outburst of XTE~J1118+480 ($\sim 10$ days), we can estimate the value of
the inner radius of the quiescent accretion disk in this system. The
characteristic viscous time scale for a gas pressure dominated
accretion disk with dominant free-free opacity \citep{SS73} is 
\begin{eqnarray}
\label{eq:tvisc}
t_{visc} \approx 3.7\times 10^{5} \alpha^{4/5} \dot{M}
_{16}^{-3/10} m_{1}^{1/4} R_{10}^{5/4} s, 
\end{eqnarray} 
where $\alpha$, $\dot{M}_{16}$, $m_{1}$ and $R_{10}$ are the disk viscosity 
parameter, disk accretion rate in units of $10^{16}$ g/s, mass of the compact 
object in solar units, and distance from the compact object in units of 
$10^{10}$ cm respectively. Here and elsewhere we use the symbols
$R$ and $r$ to denote distance in physical and Schwarzschild units
respectively ($r = R/R_{s} = R c^{2}/ 2 G M_{1}$).  The accretion rate
can be estimated using: $L \sim L_{X-ray} \sim 0.1\dot{M}c^2$.  Using an
ASM count-rate of 2.8 c/s and a distance of 1.1 kpc, the peak X-ray
luminosity should be $\sim 1.3\times10^{36} erg/s$ \citep{Fender01},
implying an accretion rate of $\dot{M} \approx 1.5 \times 10^{16}$.

Assuming the X-ray delay time, $t_{d}$ to be equal to the viscous
time, $t_{visc}$, one can calculate the initial value of the
transition radius, $R^{tr}$: 
\begin{eqnarray}
\label{eq:rtr}
R^{tr}_{10} \approx \left ( \frac{t_{d}}{3.7 \times 10^{5} s} \right )^
{4/5} \alpha^{-16/25} \dot{M}_{16}^{6/25} m_{1}^{-1/5} 
\end{eqnarray}
Using the values from the \citet{Esin01} model and our own
observations; $\alpha = 0.25$, $m_{1} = 9$, $\dot{M}_{16} = 1.5$, and
$t_{d} = 10$, one obtains $R_{10}^{tr} \sim 3.4$ and $r^{tr} \sim 1.2
\times 10^{4}$.

We can estimate the outer radius of the accretion disk, $R_{out}$ for
comparison to the transition radius calculated above.  Using numerical
integration, \citet{Eggleton83} approximates the effective radius of
the Roche lobe as:
\begin{eqnarray}
\label{eq:rl}
\frac{R_{l}}{a} \approx 1 - \frac{0.49 q^{2/3}}{0.6 q^{2/3}+ln(1+q^{1/3})}
\end{eqnarray}
Where $R_{l}$ is the radius of the primary Roche lobe, $a$ is the
binary separation, and $q=M_{2}/M_{1}$.
Using $M_{1}=9M_{\odot}$ and $M_{2}=0.5M_{\odot}$ we obtain
$R_{l}\approx 0.825 a$.  Using the orbital parameters from
\citet{McClintock01}, we have $a_{2}~sin(i) \approx
2.35\pm0.05~R_{\odot}$, where $a_{2}$ is the distance from the
secondary to the center of mass and $i$ is the inclination of the
system.  Assuming $i \approx 70^{\circ}$ and
$a=a_{2}(M_{1}+M_{2}/M_{1})$, we get $R_{l} \approx 1.5\times10^{11}$
cm.  If we then assume the outer radius of the accretion disk is some
fraction of $R_{l}$ \citep{LP79}, say 80\%, then $R_{out} \approx
1.2\times10^{11}$ cm and $r_{out} \approx 6.7 \times 10^{4}$.

While our estimated value for $r^{tr}$ of XTE~J1118+480 is similar
to that calculated for GRO~J1655--40 by \citet{Hameury97}, the outer
disk radius, $r_{out}$, is likely smaller than found in typical
X-ray nov{\ae}.  According to the DIM model, it is the outer radius
which should determine the total
mass stored in the accretion disk before the onset of the outburst.
If we assume a constant coefficient of mass conversion into emission,
the accretion disk mass then determines the integral flux emitted 
during outburst.  This model may explain qualitatively why the level
of X-ray emission detected from XTE~J1118+480 was unusually low,
whereas the temporal evolution of the outburst was similar to
other typical X-ray nov{\ae}, such as XTE~J1550--564 \citep{Jain01}.

\section{Summary}

We have presented data from the ROTSE-I telescope that show the
optical counterpart of XTE~J1118+480 during two outbursts in 
the early part of 2000.  A comparison of the ROTSE-I optical data
with the RXTE/ASM X-ray data during the second outburst show that 
XTE~J1118+480 appeared first in the optical band, approximately
7 days before appearing in X-rays.  Additionally, the optical flux
rose at a much greater rate than the X-ray flux during the initial
stages of the second outburst.  At the half-power point, the
delay was approximately 10 days.

A possible explanation for the X-ray delay is that the accretion flow
consists of two components.  The outer component is a classical
geometrically thin, optically-thick disk.  This flow then becomes
an optically-thin corona, such as an ADAF, as it moves inward toward
the central compact object.  During outburst, the radius at which this
transition occurs moves inward on a viscous time-scale, eventually
resulting in X-ray emission.  By measuring the delay between the onset
of the optical emission and the X-ray emission, the initial radius of
the transition region may be estimated.  In the case of XTE~J1118+480
we may approximate the initial transition radius by using the delay
between the ROTSE-I and ASM observations of the March
outburst.  Given a 10 day delay, we estimate the transition
radius to have been $\sim 1.2 \times 10^{4}$ Schwarzschild radii, or 
$\sim 3.4 \times 10^{5}$ km.

We would especially like to thank the RXTE/ASM team for their data
on XTE~J1118+480.  Work performed at LANL is supported by NASA SR\&T
through DOE contract W-7405-ENG-36 and through internal LDRD funding.
Work performed at University of Michigan is supported by NASA under
SR\&T grant NAG5-5101, the NSF under grants AST-9703282 and
AST-9970818, the Research Corporation, the University of Michigan,
and the Planetary Society.  Work performed at LLNL is supported by
NASA SR\&T through DOE contract W-7405-ENG-48.

\clearpage
\begin{deluxetable}{cccc}
\tablewidth{0pt}
\tablecaption{Selected ROTSE-I data on XTE J1118+480}
\tablehead{
   \colhead{Date} & 
   \colhead{MJD\tablenotemark{a}} &
   \colhead{Magnitude\tablenotemark{b}} &
   \colhead{Flux\tablenotemark{c}}
}
\startdata
12/18/99 & 51530.315 & $\downarrow$15.3 & $\downarrow$2.75 \\
12/28/99 & 51540.283 & $\downarrow$14.5 & $\downarrow$5.75 \\
\tableline
1/04/00 & 51547.223 & 13.41$\pm$.08 & 15.67$\pm$1.10 \\
1/12/00 & 51555.195 & 13.26$\pm$.02 & 18.03$\pm$0.41 \\
1/24/00 & 51567.270 & 14.00$\pm$.15 & 9.10$\pm$1.28 \\
2/02/00 & 51576.305 & 14.58$\pm$.08 & 5.35$\pm$0.38 \\
2/08/00 & 51582.148 & 15.26$\pm$.18 & 2.85$\pm$0.47 \\
\tableline
2/15/00 & 51589.313 & $\downarrow$15.0 & $\downarrow$3.63 \\
2/23/00 & 51597.184 & $\downarrow$14.8 & $\downarrow$4.37 \\
\tableline
2/26/00 & 51600.301 & 14.92$\pm$.13 & 3.95$\pm$0.41 \\
3/01/00 & 51604.246 & 13.80$\pm$.03 & 10.95$\pm$0.32 \\
3/05/00 & 51608.164 & 13.49$\pm$.02 & 14.54$\pm$0.32 \\
3/12/00 & 51615.176 & 13.28$\pm$.02 & 17.76$\pm$0.33 \\
4/20/00 & 51654.227 & 13.26$\pm$.05 & 18.07$\pm$0.87 \\
6/26/00 & 51721.195 & 13.27$\pm$.06 & 17.85$\pm$1.03 \\
\enddata
\tablenotetext{a}{Modified Julian Day (jd-2400000.5)}
\tablenotetext{b}{Unfiltered optical ccd, $\downarrow$ values indicate
sensitivity of image in which XTE J1118+480 is not detected.}
\tablenotetext{c}{$10^{-15} erg/s/cm^{2}/\textrm{\AA}$}
\end{deluxetable}

%

\clearpage

\clearpage

\end{document}